\documentclass{gGAF2e}

\begin{document}

\jvol{00} \jnum{00} \jyear{2013} 

\markboth{Dikpati}{Solar/stellar meridional circulation}


\title{{\textit{Meridional Circulation From Differential Rotation in an 
Adiabatically Stratified Solar/Stellar Convection Zone}}}

\author{M. DIKPATI\thanks{Email: dikpati@hao.ucar.edu 
}\\\vspace{6pt}  High Altitude Observatory, NCAR,
3080 Center Green Dr., Boulder, Colorado 80301, USA\\ 
\vspace{6pt}\received{June 2013 released date year} }

\maketitle

\begin{abstract}
Meridional circulation in stellar convection zones is not generally well
observed, but may be critical for the workings of MHD dynamos operating in 
these domains. Coriolis forces from differential rotation play a large role in
determining what the meridional circulation is. Here we consider the
question of whether a stellar differential rotation that is constant on 
cylinders concentric with the rotation axis can drive a meridional circulation.
Conventional wisdom says that it can not. Using two related forms of the
governing equations that respectively estimate the longitudinal components of
the curl of the meridional mass flux and the vorticity, we show that  
such differential rotation will drive a meridional flow. This is because 
to satisfy anelastic mass conservation, non-spherically symmetric pressure
contours must be present for all differential rotations, not just ones that 
depart from constancy on cylinders concentric with the rotation axis.  
Therefore the fluid is always baroclinic if differential rotation is present. 
This is because, in anelastic systems, the perturbation pressure must satisfy 
a Poisson type equation, as well as an equation of state and a thermodynamic 
equation. We support our qualitative reasoning with numerical examples, and 
show that meridional circulation is sensitive to the magnitude and form of 
departures from rotation constant on cylinders. The effect should be present
in 3D global anelastic convection simulations, particularly those for which
the differential rotation driven by global convection is nearly cylindrical
in profile. For solar-like differential rotation, Coriolis forces generally
drive a two-celled circulation in each hemisphere, with a second, reversed 
flow at high latitudes. For solar like turbulent viscosities, the 
meridional circulation produced by Coriolis forces is much larger than 
observed on the Sun. Therefore there must be at least one additional force,
probably a buoyancy force, which opposes the meridional flow to bring
its amplitude down to observed values.


\end{abstract}

\section{Introduction}

Meridional circulation has been observed at and near the surface of the Sun 
since at least the 1980's by a variety of methods, including surface Doppler 
shifts, helioseismic techniques, and feature tracking 
\citep{u10,getal06,kghbrh12,ba10,rhk12,sks06,skz07}. Results from the various 
methods, including earlier measurements, are compared in \citet{u10}. All 
results indicate that in low and mid-latitudes the meridional flow is toward 
the poles at the solar surface, but there is no concensus currently on whether 
this flow extends all the way to the poles, or is replaced by a flow toward 
the equator poleward of about $60^{\circ}$.

Meridional circulation is thought to be present in most rotating stars with 
convection zones. This circulation is of particular interest because it may 
play a critical role in solar and stellar dynamos, particularly flux-transport
dynamos \citep{ws91,csd95,d95,dc99}. For example, whether in 
the Sun the meridional flow has one cell or two cells in each hemisphere may 
determine what causes some solar cycles to be longer than others
\citep{dgdu10}. Meridional circulation also plays an important role 
in the angular momentum balance of a stellar convection zone. There too
models suggest there can be either one or two cells in latitude \citep{tt95}.

Because meridional circulation is very difficult to measure, especially below 
the surface, to estimate it requires resorting to other methods, such 
as inferring it theoretically from other better known quantities, such as 
differential rotation. There are several approaches that can be taken to the 
problem of a theory of meridional circulation for stellar convection zones. 
One is to build full 3D HD/MHD models of the convection zone, such as has 
been done with the ASH code \citep{m05} and others \citep{zsz12}, which 
generate both differential rotation and meridional circulation. A second is 
to build an axisymmetric mean field theory of coupled differential rotation 
and meridional circulation, as described in \citet{r89,r05}. A third, more 
limited approach is to infer theoretically what meridional circulation is 
most consistent with a particular specified differential rotation, which is 
more clearly known throughout the convection zone from observations. That 
is the approach we follow here. It is related to the approach defined by 
\citet{d96}.

Even using this more limited approach, there are many forces to consider
including in the theory for meridional circulation, some known from 
observations and basic theory much better than others. These forces include
Coriolis forces from differential rotation, pressure forces, turbulent
viscous forces, organized turbulence, buoyancy forces, electromagnetic body
forces, and possibly others. We can gain insight into which forces can be
responsible for which properties of meridional circulation by starting with
a very limited set, and then adding to them in succession. In this study,
we therefore limit ourselves to solutions involving the first three forces,
set aside organized turbulence and electromagnetic body forces, and 
consider the role of buoyancy forces only in qualitative reasoning. Another
reason to leave out thermodynamics initially is that including it leads
to the possibility of exciting axisymmetric convective rings in a system
in which the non axisymmetric convective turbulence is treated only in
parametric form. Such rings, if excited, would look very different from 
the observed meridional circulation; there would typically be several in
each hemisphere. In a full 3D global simulation these axisymmetric modes
are likely to be of much lower amplitude than 3D nonaxisymmetric convective
modes, though they could still be present. In any case, here we wish to
isolate driving that could be responsible for just one or two cells between
equator and poles, such as is observed.

This allows us to focus particularly on the role played by Coriolis forces 
coming from the well known differential rotation. Compressibility does enter 
the problem through the radial density gradient and its role in the equation of
mass conservation, as well as in the equations of motion. Thermodynamics, which
is bound to be important, is omitted initially because how to specify it
in a nearly adiabatically stratified convection zone is not that well known
\citep{r05}. These thermodynamic effects could include subadiabatic parts
of the convection zone postulated to arise from nonlinear convective
transport effects \citep{ss91}, a subadiabatic tachocline below the
convection zone whose thermodynamic properties 'leak' into the convection
zone and latitudinal entropy gradients from latitudinal turbulent heat 
transport caused by the latitudinally varying influence of rotation upon 
convection. These additions are reserved for later studies, once the role 
of Coriolis forces from the differential rotation has been studied.

Viewed from an inertial (nonrotating) reference frame, conventional wisdom says
that centrifugal forces arising from a differential rotation that is constant
on cylinders are conservative (can be represented as the gradient of a 
potential) and therefore can not drive a meridional circulation (see, e.g.,
\citet{rh04}, equation 3.13 and associated text). The same argument can be 
made using a rotating frame provided the centrifugal force due to the
rotation of the reference frame is incorporated into gravity; except for very
fast rotating stars, this produces very little departure from a spherical
shape. In this case it is the Coriolis forces from cylindrical differential 
rotation that are conservative. It is important to recognize that this
conventional wisdom is usually inferred from the equation for the azimuthal
component of vorticity, that is, the equation that calculates the vorticity
of the meridional flow itself. In this equation the Coriolis terms cancel
when the differential rotation is constant on cylinders, a consequence of
the Coriolis forces per unit mass being conservative.

While it is true that centrifugal or Coriolis forces per unit mass arising 
from a cylindrical differential rotation are conservative, it is 
equally true that these forces, written as forces per unit volume, are 
not. The equations of motion per unit mass and per unit volume each
can be used to calculate meridional circulation from Coriolis and/or
other forces. Both approaches should lead to the same meridional
circulation for a given differential rotation. In what follows below, we will
first develop equations for meridional circulation from the equations of 
motion per unit volume, by taking the curl of the meridional mass flux, 
arriving at an equation analogous to equation (47) of \citet{r89}.
We will solve this equation for meridional circulation streamlines 'driven'
by Coriolis forces due to differential rotation. Our primary focus in this
paper will be on differential rotation that is constant on cylinders. We will
then show that our results should be compatible with those obtained by solving
for streamlines from the equation for the longitudinal component of vorticity,
which is derived from the equations of motion per unit mass. The key to 
understanding the compatibility of these approaches is the differing roles
played by the fluid pressure in the two systems.

In the Sun, thermodynamic surfaces would be almost spherical, since 
centrifugal effects are five orders of magnitude smaller than Newtonian 
gravity. So it is very tempting to conclude that in an adiabatically 
stratified rotating convection zone meridional circulation will be driven 
by only the part of the centrifugal or Coriolis forces that arise from 
departures of the differential rotation from constancy on cylinders. 
Is it certain that in the presence of rotation, an adiabatically 
stratified convection zone truly will have all thermodynamic surfaces coincide? 
We already know that if we add a magnetic field to an adiabatic layer, we 
can get density variations and therefore departures from coincidence of 
the thermodynamic surfaces -- the well established phenomena of magnetic 
buoyancy. Could the presence of rotation do something analogous to the
thermodynamics? 

Results from both mean-field and full 3D simulations of differential rotation
and meridional circulation also relate to this issue. As discussed in 
\citet{rh04} just below equation (3.13) and elsewhere, there are 
numerous numerical simulations of differential rotation and meridional 
circulation driven by global convection in a rotating spherical shell 
in which, in the bulk of the convecting shell, the rotation profile is very close to 
cylindrical, but induced meridional circulation is present. This is true for
both the incompressible and compressible (anelastic) cases. This circulation is 
not confined to boundary layers where we might expect deviations from the 
cylindrical rotation profile to occur. 

In addition, the argument is commonly made for both mean-field and full 3D
convective models of differential rotation and meridional circulation,
e.g. \citet{r05}, that it is the outward radial Coriolis force due to the 
'equatorial acceleration' built up by equatorward angular momentum transport 
by turbulent Reynolds stresses that is responsible for the fact that the 
induced meridional flow is outward near the equator, and therefore poleward
near the outer boundary of the convection zone. To be valid, this argument 
should not depend on just that part of the differential rotation that
deviates from constancy on cylinders. This reasoning is also independent of
the thermodynamics of the system. In other words, it does not depend on
the meridional circulation being directly thermally driven by buoyancy forces
acting in the meridional plane.

\section{Equations}

The equations of motion we start from can be written either in an inertial or
a rotating reference frame. We prefer the rotating system, because that is 
virtually always the system used for numerical simulations. Within this system, 
we assume that the centrifugal force per unit mass of the rotation is absorbed
into the gravitational potential, as is customarily done for slow rotators 
like the Sun, and treat the system as if it is truly spherical, even though in
reality it deviates from that by a few parts in $10^5$. 

In a radiative zone of a star, this deviation would lead to the so-called 
Eddington-Sweet currents, but these are not relevant to stellar convection 
zones where the meridional circulation induced by the turbulent convection 
and the differential rotation driven by it will be much larger. We also 
subtract out a spherically symmetric hydrostatic balance, and assume there 
are no imposed latitudinal gradients of density or pressure, so the reference 
state density $\rho_o$ and pressure $p_o$ are themselves functions
only of the radial coordinate $r$. We then use $\rho,p$ for the density
and pressure departures from their reference state values at the same point 
in the domain. Thus, $\rho$ is the Eulerian density, which in a nearly
adiabatically stratified stellar convection zone is always much smaller than
the reference state density $\rho_o$ at the same place in the spherical shell.
Even though we are focussing here primarily on a mechanical model without
thermodynamics, we retain the perturbation density when coupled with gravity
in order to make certain qualitative arguments about the equations.
We also ignore inertial nonlinearities (terms containing products of 
velocities), because they are not critical to the reasoning or the results, 
and because in reality the observed meridional circulation in the Sun is 
relatively small, $10-20\,{\rm m}{\rm s}^{-1}$ at most. Furthermore, we include in
the model only the parts of the turbulent Reynolds stresses that can be
represented by a turbulent viscosity, and we take this viscosity to be a 
specified function of radius.

With these conditions, and defining the velocity components in the standard
spherical coordinates $r,\theta,\phi$ as $w,v,u$, we can write the equations 
of motion per unit volume in the meridional plane as

$$\rho_o{\partial w \over \partial t}=-{\partial p \over\partial r}
+2\Omega\sin\theta\rho_o u-\rho g+\rho_oV_r; \quad\eqno(1)$$
 
$$\rho_o{\partial v \over \partial t}=-{1 \over r}{\partial p \over
\partial \theta}+2\Omega\cos\theta\rho_o u +\rho_oV_{\theta}.\quad\eqno(2)$$
 
Note that in the anelastic approximation, the Eulerian perturbation density
$\rho$ does not enter the mass continuity equation to lowest order, whether
or not the domain is adiabatically stratified. In equations (1) and (2), $u$ 
is the linear rotational velocity relative to the rotating frame, which can 
be written as $u=r\sin\theta\omega$, in which $\omega$ is the local angular 
velocity relative to the same frame. This frame rotates at rate $\Omega$, 
which for a convection zone would be conveniently taken as the rotation rate 
of the interior below it. $g$ is the local gravity, which itself can be a 
function of $r$. Finally, $V_r,V_{\theta}$ represent the turbulent viscous 
forces per unit mass associated with an isotropic turbulent viscosity, which 
itself could be a function of radius. For brevity, we  keep these forces 
in symbolic form.

Since all these velocities will be small compared to the sound speed, 
they are constrained by the mass continuity equation (the anelastic system,
a very good approximation for stellar convection zones)

$${1 \over r}{\partial \over \partial r}(\rho_o r^2 w)+{1 \over 
\sin\theta} {\partial \over \partial \theta}(\rho_o \sin \theta v)=0.
\quad\eqno(3)$$

As in many applications it is useful to define a streamfunction $\chi$ for 
the meridional velocities $w,v$ such that

$$\rho_o w={1 \over r \sin \theta}{\partial \over \partial \theta}
( \sin \theta \chi); \rho_o v=-{1 \over r}{\partial \over \partial r}
(r \chi),\quad\eqno(4)$$

\noindent so that equation (3) is satisfied identically.

Then we can derive an equation for the curl of the meridional mass flux, from
which we can compute the circulation streamlines represented by $\chi$. We
eliminate the perturbation pressure from equations (1) and (2) by 
differentiating  equation (2) with respect to $r$, (1) with respect to 
$\theta$, and subtracting equation (1) from equation (2); the resulting 
equation for $\chi$ can be written as 

$${\partial \over \partial t}L \chi=-{2\Omega \cos \theta \over r}{\partial
\over \partial r}(r\rho_o u)+{2\Omega \over r} {\partial \over \partial 
\theta}(\sin \theta \rho_o u)-{g(r) \over r}{\partial \rho \over 
\partial \theta}-M(\chi),\quad\eqno(5)$$

in which $L =\nabla^2 ()-()/r^2\sin^2\theta$ ($\nabla^2$ 
the axisymmetric Laplacian operator) and $M(\chi)$ is a fourth order operator 
acting on $\chi$ that contains all of the viscous Reynolds stress terms.

Then if we want to find steady state solutions for the meridional 
streamfunction in this system, we bring the term $M(\chi)$ to the left hand 
side, which yields

$$M(\chi)=-{2\Omega \cos \theta \over r}{\partial \over \partial r}(r
\rho_o u)+{2\Omega \over r} {\partial \over \partial \theta}(\sin \theta 
\rho_o u)-{g(r) \over r}{\partial \rho \over \partial \theta}.\quad\eqno(6)$$

Each term in equation (6) can be identified with a term in equation (47) of
\citet{r89}. Equation (6) says there is meridional circulation driven by 
Coriolis forces unless the Coriolis terms cancel, and/or by density 
perturbations if there are any. By inspection we see that the Coriolis terms
do not cancel regardless of the profile of the reference density $\rho_o$,
so long as it is not constant. It is also noteworthy that equation (6)
contains no term directly involving the perturbation pressure, so it appears
that any baroclinic effect must enter through the perturbation density.
We will show below that a baroclinic effect can also be produced from the 
reference state density coupled with the Coriolis forces, even if there is
no gravity to produce buoyancy forces. This is because these Coriolis forces
are closely linked to the perturbation pressure. 

We can find that link by forming an equation for the divergence of 
the mass flux by the meridional flow, found by taking the divergence in the
meridian plane of equations (1) and (2). From the mass continuity equation, 
equation (3), this divergence must be zero, even if the solution is time 
dependent. This yields the following relationship between pressure
and Coriolis forces, plus whatever other forces are present, in this case 
viscous forces and perhaps buoyancy. The result is a Poisson type equation
for the pressure, given by

$$\nabla^2 p=2\Omega\left({1 \over r^2}{\partial \over \partial r}
(\rho_or^3\sin^2\theta\omega)+{1 \over \sin\theta}{\partial \over \partial 
\theta}(\sin^2\theta\cos\theta\rho_o\omega)\right) -{1 \over r^2}{\partial
\over \partial r}(r^2g\rho)$$
$$+{1 \over r^2}{\partial \over \partial r}(r^2 \rho_o V_r)+{1 \over \sin
\theta}{\partial \over \partial \theta} (\sin \theta\rho_oV_{\theta}).
\quad\eqno(7)$$

All numerical models for anelastic systems must take account of this 
relationship to find solutions. From equation (7) with a reference state 
radial density gradient, it is clear that the Coriolis terms contribute to 
a nonzero perturbation pressure $p$ even if the density perturbation $\rho$ 
vanishes everywhere, as is commonly assumed for nonmagnetic adiabatically 
stratified medium. Only if the reference state density is constant and 
the relative angular velocity is constant on cylinders does the Coriolis 
'source' for pressure perturbations vanish. This is because there are 
two distinctly different roles that density plays in the equations of motion.
One is in generating buoyancy forces, when the density of a fluid element 
in a gravity field differs from that of its surroundings. The second role 
is one of inertia. Even without a gravity field present, variations of density
with position, such as a radial gradient, can cause dynamical effects which 
will generally be associated with perturbation pressure variations.

What forcing function do we get in equation (7) when we take differential 
rotation to be constant on cylinders and we temporarily ignore the viscous 
and buoyancy forcing terms there? Does any forcing function remain? It must,
if Coriolis forces are to produce a pressure perturbation in the absence of 
any other forces. It is straightforward to demonstrate that under these 
conditions, the Coriolis forcing terms in equation (7) reduce to

$$2\Omega\rho_o\left(\left(1+{r \over H_{\rho}}\right)\omega+s{\partial
\omega \over \partial s}\right).\quad\eqno(8)$$
in which $s=r\sin\theta$ is the coordinate perpendicular to the rotation
axis. In the special case when $\omega$=constant, then $\partial \omega 
/\partial s=0$ and the remaining constant rotation $\omega$ term can be 
incorporated into the gravitational potential, resulting in no pressure 
perturbation relative to a modified rotating frame that rotates at the rate 
$\Omega+\omega$. But when $\partial \omega /\partial s\not=0$ so $\omega$ is 
not constant, in general a pressure perturbation will result from expression 
(8) and the system remains baroclinic, even though the differential rotation
is still constant on cylinders.

Given the reasoning above, it is instructive to consider a hierarchy of 
cases of increasing physical complexity for which the differential rotation 
is constant on cylinders.

With constant density, no thermodynamics and no equation of state,
there will be no meridional circulation, because the Coriolis force
per unit mass and per unit volume are both conservative. The forcing
of pressure by Coriolis forces is absent in equation (7).

If a radial density gradient is added, but still no thermodynamics or 
equation of state, there will be meridional circulation for a cylindrical 
differential rotation profile case, because Coriolis forcing is present 
in equation (7), since now the Coriolis force per unit volume is no longer
conservative. This leads to departures of the pressure field from 
spherically symmetric, so the pressure and density surfaces no longer
coincide; the fluid is baroclinic, even without gravity or thermodynamics.

If there is no radial density gradient associated with Coriolis forces, 
but there are density perturbations and gravity present (the Boussinesq 
system), there will be meridional circulation. Here the Coriolis forcing
vanishes in equation (7), but there is buoyancy forcing, so again
the pressure and density surfaces do not coincide, and the system is
baroclinic. In this case the meridional circulation arises from buoyancy
effects from the radial density gradient, not inertial effects.

With a radial density gradient, gravity, a thermodynamic equation and 
an equation of state included (the anelastic system), there will be 
meridional circulation when there is cylindrical differential rotation.
Here both the inertial and the buoyancy effects of the radial
density gradient are at work.

Finally, for each of the cases described above, if the differential 
rotation present is not constant on cylinders, Coriolis forces
from differential rotation will drive a meridional circulation.
For these cases, the Coriolis forcing in equation (7) does not vanish.

\section{Meridional circulation from differential rotation: examples}

Here we show that from a purely mechanical model for meridional circulation,
by solving equation (6) with no buoyancy forces included, cylindrical
differential rotation drives a meridional circulation when a radial density
gradient is present. We use so-called multi-grid methods, and apply 
boundary conditions on the streamfunction that require there be no
meridional flow into or out of the domain (so the boundaries are a streamline)
and there is no viscous stress associated with the meridional flow
at the top and bottom.  The turbulent viscosity profile is given by

$$\nu(r)=\nu_c+{\nu_s \over 2}\left(1+erf\left({2(r-r_c) \over d_{\nu}}\right)
\right),\quad\eqno(9)$$

in which $r_c=0.7R_{\odot}$, $d_{\nu}=0.1R_{\odot}$ and for solar viscosities
we take $\nu_c=7 \times 10^{11}\,{\rm cm}^2{\rm s}^{-1}$ and $\nu_s=6 \times
10^{14}\,{\rm cm}^2{\rm s}^{-1}$. This profile is relatively independent 
of $r$ through most of the convection zone, declining to a substantially 
lower value near the bottom of the convection zone and below, where the 
tachocline is found. 

If we were solving for differential rotation instead of specifying it, we
would need to consider the role of the 'non-viscous' part of the turbulent 
Reynolds stresses, since they transport angular momentum in latitude 
\citep{r05}. They do not enter directly into the equations for meridional flow,
but rather their effects are implicit in the form of the differential rotation
taken, since the differential rotation results from the effects of viscous
and nonviscous parts of the Reynolds stress together with Coriolis forces
from the meridional flow. One could add to the meridional flow equations
forcing terms from the part of the Reynolds stress that represents nonviscous
transports of angular momentum in the meridional plane, coming from the
correlation of turbulent meridional flow components with each other, in
particular, $v'v'$, $w'w'$ and $v'w'$. But little is known about these 
quantities in stellar convection zones, so we leave them out of our analysis.

To be specific in the Coriolis forcing, we fit a cylindrical differential 
rotation to the observed solar surface differential rotation.  Early 3D global
convection models for the Sun \citep{gm86,m05} produced differential rotations 
that were solar-like near the surface, but nearly cylindrical in profile 
beneath. For the density stratification, we use a polytropic form given by 

$$\rho(r) = \rho_b\left(\frac{R}{r} - 0.97\right)^m.\quad\eqno(10)$$ 

If the ratio of specific heats $\gamma=5/3$ as in a monatomic gas, the case 
$m=1.5$ corresponds to an adiabatically stratified convection zone. For
convenience in these calculations, we simply increase $m$ from zero to 1.5
to illustrate the effect of increasing the radial density gradient on the
pattern of meridional circulation found, without identifying different
$m$ values with different polytropes formally. The cylindrical differential 
rotation we use is given by
$$\Omega(r,\theta)=C_1+C_2 {\left({r\sin\theta \over R}\right)}^2
- C_3 {\left({r\sin\theta \over R} \right)}^4, \quad\eqno(11a)$$
in which $C_1=\Omega_{\rm Eq} - \Omega_{\rm c} - a_2 - a_4$, $C_2=
a_2+2 a_4$ and $C_3=a_4$. The values of $\Omega_{\rm Eq}=460.7 \times 2\pi$, 
$\Omega_{\rm c} =432.8 \times 2\pi$, $a_2=-62.9 \times 2\pi$ and $a_4=
-67.13 \times 2\pi$, all in nHz units, are matched to the solar surface 
differential rotation. Figure 1 shows a plot of this differential rotation.

\setcounter{figure}{0}
\begin{figure}
\includegraphics[width=15.5cm]{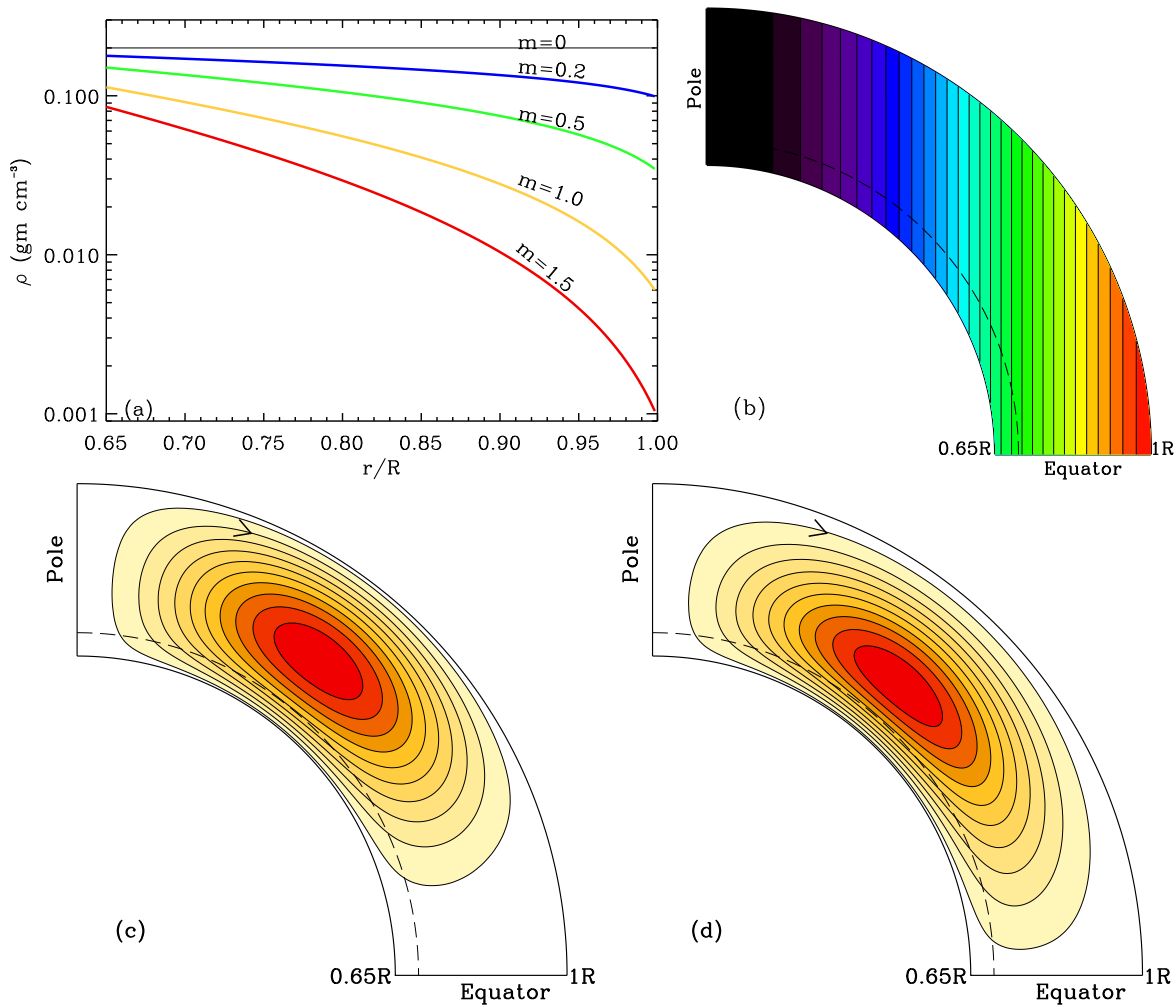}%
\caption{Meridional circulations driven by solar surface differential rotation
matched to a cylindrical profile (Panel (b)) for radial density profiles shown
in Panel (a). Circulation in Panel (c) is for polytropic index m=0.2, Panel (d)
for m=1.5 (adiabatic convection zone)}
\label{cyldr}
\end{figure}

Figure 1 displays the density (panel (a)) and differential rotation (panel (b))
profiles, as well as two resulting meridional circulations (panels (c),(d)).
We see that for both weak and strong density gradient, we get a single-celled
meridional flow, but one that has equatorward flow near the outer boundary. This
is opposite to what is observed on the Sun, but reminiscent of what \citet{k70}
found with his mechanical model. There the mechanism is different, because he
assumed an anisotropic eddy viscosity. 

As we should expect, the amplitude of the meridional circulation driven by
a particular differential rotation amplitude is inversely proportional to
the turbulent viscosity used in our linear model. For a turbulent viscosity 
of $10^{14}\,{\rm cm}^2{\rm s}^{-1}$, at least an order of magnitude larger 
than commonly estimated for the solar convection zone, peak meridional circulation 
amplitudes are $\sim 200\,{\rm m}{\rm s}^{-1}$, an order of magnitude larger 
than observed in the Sun. It takes a turbulent viscosity of $10^{15}\,
{\rm cm}^2{\rm s}^{-1}$, two orders of magnitude larger than solar values, 
to bring the velocities down to observed values. 

We can draw two immediate conclusions from these results. First, the solar
differential rotation profile must depart significantly from cylindrical in
order to get the correct sense of circulation (poleward near the solar 
surface). Second, there must be at least one additional force present that
opposes the driving effect of Coriolis forces from the differential rotation, 
to bring the circulation amplitudes down to solar values. Our opinion is that
this force is most likely to be a buoyancy force arising from one or more
of the thermodynamic effects listed in the introduction. The problem is that,
as stated in the introduction, exactly what form these effects might take
is unknown from observations, and also not well developed in theory. They
would all cause the convection zone thermodynamic surfaces to depart from 
spherical. Conventional wisdom might say that they should take the form
that follows from assuming there is a 'thermal wind' balance in the turbulent
convection zone with no meridional circulation at all. But we have
demonstrated above that, because of the radial density gradient, this is not 
true even when differential rotation is constant on cylinders.

What happens to the meridional circulation pattern when we modify the
differential rotation profile to depart from cylindrical, even slightly? 
To examine this question, we consider a purely latitudinal variation
in the bulk of the convection zone, and convergence to a single intermediate
rotation rate at the bottom, thereby including a form of tachocline there
(see expression 11b). This profile is given by,

$$\Omega(r,\theta)=\Omega_{\rm c}+{1 \over 2}\left[1+{\rm erf}\left(2 {r-r_c
\over d_1}\right)\right]{\Omega_{\rm s}(\theta)-\Omega_{\rm c}},
\quad\quad\eqno(11b)$$
in which, $\Omega_{\rm s}(\theta)=\Omega_{\rm Eq} + a_2 \cos^2\theta +
a_4 \cos^4\theta$ is the surface latitudinal differential rotation,
$\Omega_{\rm c}=432.8 \times 2\pi$, $\Omega_{\rm Eq}=460.7 \times
2\pi$, $a_2=-62.69 \times 2\pi$ and $a_4=-67.13 \times 2\pi$ nHz.
This profile has been widely used in solar dynamo simulations 
\citep{dc99, jetal08}. 

Then we take the approach of taking a differential rotation profile that is a 
weighted sum of a purely cylindrical profile as given in the expression (11a), 
and a purely latitudinal profile with a tachocline (as presented in the
expression 11b). 
The weighting is such that, at the outer boundary, the sum of the two profiles
add up to the observed solar profile. This combination has the fortunate 
effect of creating rotation contours that are tilted toward the poles from 
the local outward radial direction, which is a well measured characteristic
of solar differential rotation. 

In Figure 2 we display  differential rotation profiles for which (Frame a)
the total differential rotation is $70\%$ from the latitudinal profile 
and $30\%$ from the cylindrical profile, and (Frame c), the reverse of that.  
We compute the meridional circulations for these two cases, with $m=1.5$,
and display the resulting streamlines in Frames (b) and (d), respectively.
The $70\%$ case is much closer to solar observations than is the $30\%$ case.

\begin{figure}
\includegraphics[width=15.5cm]{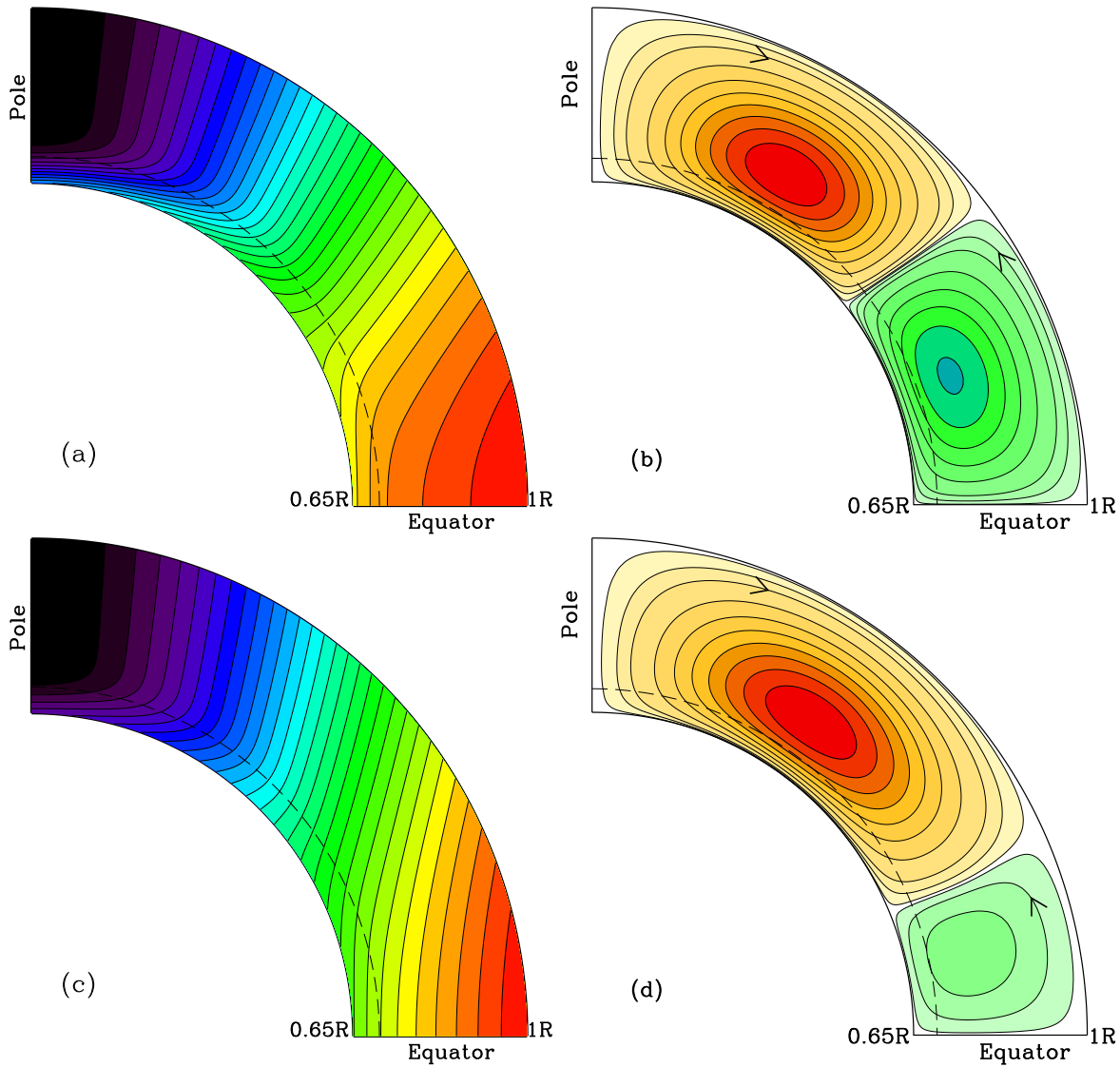}%
\caption{Mixed latitudinal and cylindrical Differential rotation and resulting
meridional circulation. Panel (a): differential rotation that is $70\%$ 
latitudinal and $30\% $ cylindrical; panel (b), the resulting meridional flow. 
Panel (c) differential rotation that is $30\%$ latitudinal and $70\%$ 
cylindrical; panel (d), the resulting meridional flow}
\label{cyldr}
\end{figure}

We see that even a relatively modest change from differential rotation 
constant on cylinders can have a profound effect. As the percentage of
the differential rotation that is independent of radius is increased, the
single meridional flow cell with equatorward flow near the outer boundary
gives way to a two-celled flow pattern, in which low and midlatitude 
surface flow is towards the poles. The higher the fraction
that is purely latitudinal differential rotation, the higher the latitude of
the boundary between the two cells. Even if the differential rotation is
taken to be totally independent of radius, the two cells remain. This 
pattern is much more in agreement with solar surface observations, such as
given in \citet{u10}. It is well established from helioseismic measurements
that the solar differential rotation does depart significantly from a
cylindrical profile \citep{h09}. As for the purely cylindrical case,
for a solar-like turbulent viscosity, the flow speeds are much larger than 
observed. So for non-cylindrical rotation profiles, the mechanism of
driving meridional circulation from differential rotation is still very
strong, requiring at least one other force to oppose the flow to bring the
speeds down to observed values.

It turns out that departures from cylindrical rotation are not the only way
a second meridional circulation cell can be introduced. If we increase the 
rotation rate of the system, but keep the differential rotation purely
cylindrical with solar-like amplitude, we get a similar effect. This is 
seen in Figure 3, which compares the solar case (Frame (a)), with one for 
$2 \times$ solar (Frame (b)), $3 \times$ solar (Frame (c)) and $5 \times$
solar (Frame (d). Keeping the differential part of the rotation the same 
in all cases means that for $2 \times$ solar the differential rotation is 
$15\%$ of the core rotation, $3 \times$ solar $10\%$ and $5 \times$ solar 
$6\%$. We see that as the basic rotation is increased from solar-like
value, a second poleward cell develops in low latitudes, which spreads 
poleward as rotation is increased, filling all latitudes by $5 \times$ 
solar. Thus the effect of the radial density gradient coupled with 
Coriolis forces can produce a variety of meridional circulations for 
cylindrical rotations, depending on the basic rotation of the system. It 
is true that the meridional circulation found is the same for different 
rotation rates but the same percentage differential rotation. The velocities 
simply scale up or down, but the profiles are the same.

\begin{figure}
\includegraphics[width=15.5cm]{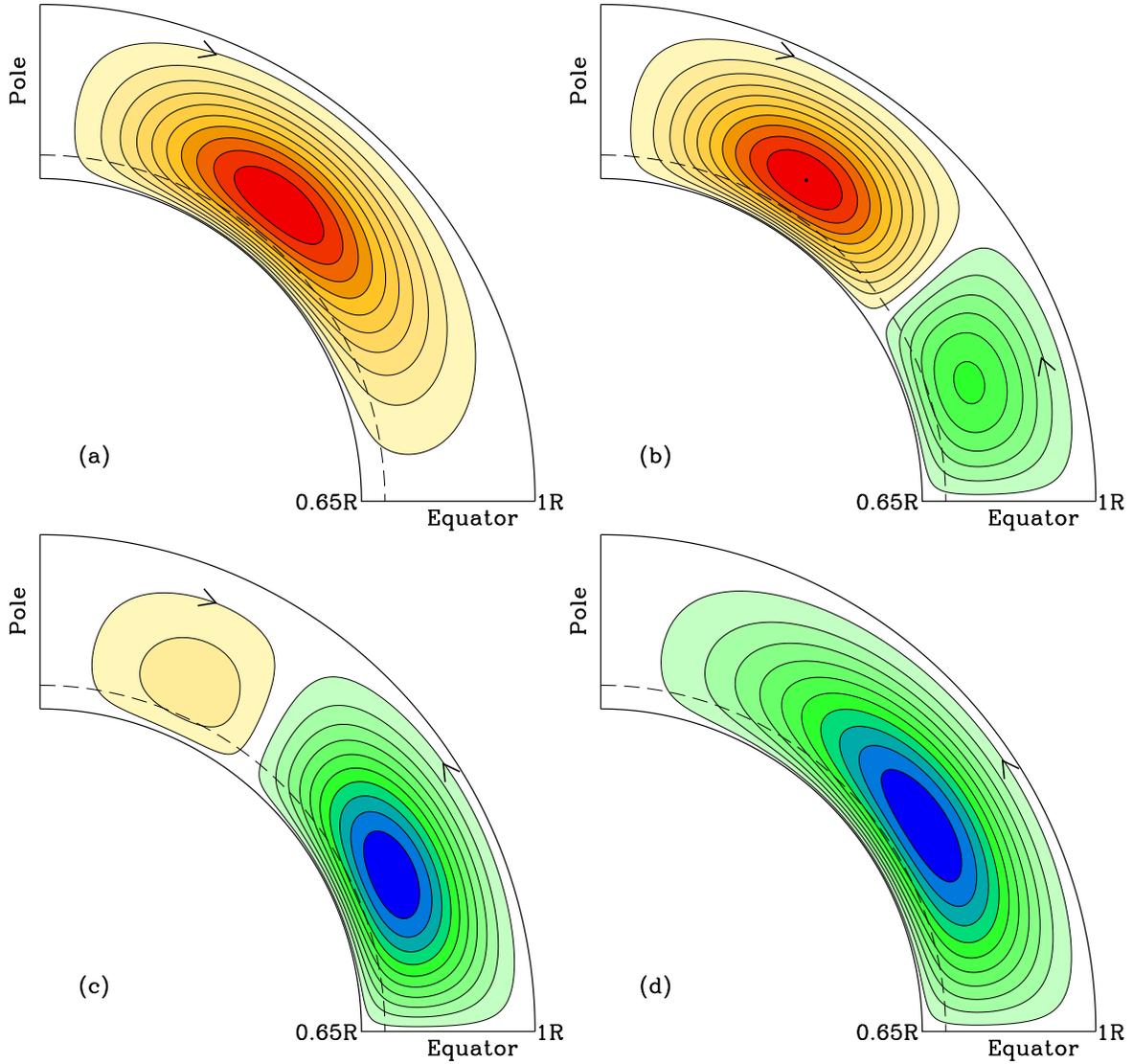}%
\caption{Meridional circulation streamlines for four rotation rates but the
same dimensional differential rotation as in the Sun. panel (a): solar; panel
(b): $2 \times$ solar; panel (c): $3 \times$ solar; panel (d): $5 \times$
solar.}
\label{cyldr}
\end{figure}

These results show that it is important to know the amount of differential
rotation in a star, as well as the basic rotation, if we are to estimate the
form of meridional circulation likely to be driven by Coriolis forces from 
the differential rotation of that star. In the stellar case, we have some
hope from asteroseismology of knowing something about differential rotation
below the stellar photosphere, but very little hope of knowing the
thermodymanics from observations.

\section{Compatability with meridional circulation from vorticity equation}

As stated in the introduction, the conventional wisdom that there will be 
no meridional circulation driven by Coriolis forces when the differential
rotation is constant on cylinders, is derived by reasoning from the vorticity
equation for the meridional circulation. It is therefore important to
examine the differences between this vorticity equation and the equation for
the curl of the meridional mass flux that we have used above.

We derive the equation for the longitudinal component of vorticity by 
dividing equations (1) and (2) by $\rho_o$,  operating on equation (2) with
$(1/r) \partial ()/\partial r$, on equation (1) with 
$-(1/r)\partial ()/\partial \theta$ and adding, to form an equation for
the $\zeta_{\phi}$, the $\phi$ component of the vorticity, which is

$${\partial \zeta_{\phi} \over \partial t}={2\Omega\cos\theta \over r}
{\partial (ru) \over
\partial r}-{2\Omega \over r}{\partial (\sin\theta u) \over \partial \theta}
$$
$$
+{1 \over r\rho^2_o}{\partial \rho_o \over \partial r}{\partial p
\over \partial \theta}+{g \over r\rho_o}{\partial \rho \over \partial 
\theta}+{1 \over r}\left({\partial V_r \over \partial \theta}-{\partial (rV_
{\theta}) \over \partial r}\right).\quad\eqno(12)$$

Then for a steady state, we bring the viscous terms to the left hand side and
have

$${1 \over r}\left({\partial V_r \over \partial \theta}-{\partial (rV_{\theta})
\over \partial r}\right)= {2\Omega\cos\theta \over r}{\partial (ru) \over 
\partial r}-{2\Omega \over r}{\partial (\sin\theta u) \over \partial \theta}
+{1 \over r \rho^2_o}{\partial \rho_o \over \partial r}{\partial 
p \over \partial \theta}+{g \over r \rho_o}{\partial \rho \over \partial 
\theta}.\quad\eqno(13)$$

We can substitute for the meridional velocities $v,w$ in terms of the 
streamfunction $\chi$ in equation (13), leading to an equation for the
streamlines that are 'forced' by Coriolis forces associated with the 
specified differential rotation, plus other forces, if any are present.
We can identify each term in equation (13) with a corresponding term in
equation (3.13) in \citet{rh04}. They are in different form here because of
subtracting out the reference state spherically symmetric hydrostatic balance.

It is readily verified that, if the relative angular velocity $\omega$ is
constant on cylinders (a function of $r\sin\theta$ only), the Coriolis terms
in equation (13) vanish, and there is no direct forcing of meridional 
circulation by Coriolis forces, in agreement with equation (3.13) of 
\citet{rh04}. This is the source of the conventional wisdom that if,
independently, the thermodynamic surfaces coincide, the terms in equation (13)
involving perturbation pressure and density will also cancel, leaving
no forcing terms for the vorticity. Hence, the conclusion that the meridional
circulation will vanish. But we have already shown in prior sections that
when there is a radial density gradient, the Coriolis forces from differential
rotation cause the density and perturbation pressure surfaces to not coincide,
so the system is baroclinic. Therefore the pressure and density terms on
the right hand side of equation (13) will not cancel, and a meridional 
circulation will be driven by them.  This is where the reasoning behind
the conventional wisdom fails, and therefore why we make a point of comparing
the two equations for calculating the meridional circulation.

In the mechanical cases shown above, where there is no gravity, it is the 
pressure term on the right hand side of equation (13) that is nonzero. Thus, 
there is, at least qualitatively, no conflict between the two approaches to
computing streamlines of meridional flow. This is because the pressure 
equation (7) applies in both approaches. Both approaches say there will 
be a meridional circulation in the compressible case with cylindrical 
differential rotation.

\section{Conclusions}

Our primary conclusion is that in an adiabatically stratified convection zone, 
the presence of any differential rotation, including one constant on 
cylinders, implies that the system is baroclinic. That is, it is one in which
the thermodynamic surfaces do not coincide. Therefore, there will be
a meridional circulation. This conclusion might appear to be somewhat in
conflict with discussion on rotating stars in books such as \citet{r89,t00} and
\citet{rh04}, perhaps even a paradox,  but the conflict or paradox is removed
if one accepts that differentially rotating adiabatically stratified 
convection zones are always baroclinic. \citet{e08,bs12} have considered some 
closely related issues, involving the differing roles played by the fluid
density in convection zones.

We also have shown that the driving of meridional circulation by Coriolis
forces from differential rotation is an extremely powerful effect, producing
flow speeds much larger than observed. It is clear that whatever additional
forces are present in the Sun must be opposing the meridional flow, to
bring speeds down to observed values when solar-like turbulent viscosities 
are used. This means that whatever thermodynamic effects are present that we
have not included must create buoyancy forces that everywhere oppose the flow.
This means that the meridional circulation can not in any sense be 
convectively driven. If they were, that would just add to the amplitude that
is already orders of magnitude too large.

For example, a single-celled meridional circulation with upward
flow near the equator and downward flow near the poles must be experiencing
a downward buoyancy force in low latitudes and an upward buoyancy force
at high latitudes. This implies that the thermodynamic structure of the
convection zone must depart from a purely spherically symmetric form into
one in which, at most depths, the fluid must be more dense near the equator
than it is near the poles at the same depth. To a first approximation, this
would imply a colder temperature at low latitudes than at high at the same
depth. This temperature structure, in turn, must be closely connected to
the tendency of the differential rotation to be in 'thermal wind' balance.

Therefore, a first approach to including thermodynamics in the mechanical 
model could be to calculate what temperature structure must
be present if the differential rotation is in thermal wind balance.
Thus, we may be able to 'reverse engineer' an estimate of the buoyancy
force to include in the mechanical model, without any thermodynamic
measurements to guide us. This approach will be explored in a future paper.
It will be important to see whether the inertial effects of the radial density 
gradient in producing meridional circulation, even when the rotation is 
constant on cylinders, are compatible with the buoyancy effects needed to
brake that meridional circulation.

The different paths to computing streamlines (from the curl of the
meridional mass flux, and from the longitudinal component of vorticity) apply 
to the same physical system, so they should yield the same circulation 
patterns. We have not attempted this comparison yet, and we are not aware 
that anyone else has either. It needs to be done. The two paths to computing 
meridional streamlines also are available in the time dependent case. In that 
case equation (5) replaces equation (6), and equation (12) replaces equation
(13), while equation (7) remains valid unchanged. 

Beyond this comparison test, there are other important questions for the 
future. For example, what form do the thermodynamic surfaces take in an 
adiabatically stratified convection zone when there is a cylindrical differential
rotation? Are these effects present in 3D numerical solutions to
global convection in rotating compressible convection zones, particularly 
those that generate differential rotation that is nearly constant on cylinders?
We hope that our results will stimulate more research on the global rotational 
dynamics and thermodynamics of stellar convection zone, with particular focus 
on theory of meridional circulation, so important to solar and stellar 
dynamo models.

\section{Acknowledgements:} 
We thank Peter Gilman for many helpful discussions on the general topic
of the Sun's global flows. We extend our thanks to two anonymous reviewers
for important comments, which have enormously helped improve the manuscript.
This work is partially supported by NASA's LWS grant with 
award number NNX08AQ34G. The National Center for Atmospheric Research is 
sponsored by the National Science Foundation.

{}

\begin{thebibliography}{9}

\bibitem[\protect\citeauthoryear{{Balbus} \& {Schaan}}{2012}]{bs12}
Balbus, S.A., Schaan, E., The stability of stratified, rotating systems 
and the generation of vorticity in the Sun, {\itshape Mon. Not. of Royal
Astron. Soc.}, 2012, {\bfseries 426}, 1546-1557

\bibitem[\protect\citeauthoryear{{Basu} \& {Antia}}{2010}]{ba10}
Basu, S. \& Antia, H.M., 2010, {\itshape The Astrophys. J.}, Characteristics
of Meridional Flows During Solar Cycle 23, {\bfseries 717}, 488-495

\bibitem[\protect\citeauthoryear{{Choudhuri}, {Sch\"ussler} \& {Dikpati}}{1995}]
{csd95} Choudhuri, A.R., Sch\"ussler, M. \& Dikpati, M.,
{\itshape Astron. \& Astrophys.}, The Solar Dynamo with Meridional Circulation,
1995, {\bfseries 303}, L29--L32
  
\bibitem[\protect\citeauthoryear{{Dikpati} \& {Charbonneau}}{1999}]{dc99}
Dikpati, M., and Charbonneau, P., {\itshape The Astrophys. J.}, 
A Babcock-Leighton Flux Transport Dynamo with Solar-like Differential Rotation,
1999, {\bfseries 518}, 508--520  

\bibitem[\protect\citeauthoryear{Dikpati et al}{2010}]{dgdu10}
Dikpati, M., Gilman, P.A., de Toma, G., and Ulrich, R.K., 
{\itshape Geophys. Res. Lett.}, Impact of Changes in the Sun's Conveyor-belt
on Recent Solar Cycles, 2010, {\bfseries 37}, L14107 1--6

\bibitem[\protect\citeauthoryear{Durney} {1995}]{d95}
{Durney}, B.R., {\itshape  Solar Phys.}, On a Babcock-Leighton Dynamo Model 
with a Deep-Seated Generating Layer for the Toroidal Magnetic Field,
1995, {\bfseries 160}, 213--235

\bibitem[\protect\citeauthoryear{Durney} {1996}] {d96} 
{Durney}, B.R., {\itshape Solar Phys.}, On the Influence of Gradients in 
the Angular Velocity on the Solar Meridional Motions, 1996, {\bfseries 169}, 
1--32

\bibitem[\protect\citeauthoryear{Evonuk} {2008}]{e08}
Evonuk, M., {\itshape The Astrophys. J.}, The Role of Density Stratification 
in Generating Zonal Flow Structures in a Rotating Fluid, 2008, 
{\bfseries 673}, 1154--1159

\bibitem[\protect\citeauthoryear{{Gilman} \& {Miller}} {1986}] {gm86}
Gilman, P.A., and Miller, J., {\itshape The Astrophys. J. Suppl.},
 Nonlinear convection of a compressible fluid in a rotating spherical 
shell, 1986, {\bfseries 61}, 585--608

\bibitem[\protect\citeauthoryear{{Gonzalez-Hernandez et al}} {2006}] {getal06}
Gonzalez-Hernandez, I., Komm, R., Hill, F., Howe, R.. Corbard, T., and Haber, 
D.A., {\itshape The Astrophys. J.}, Meridional Circulation Variability from 
Large-Aperture Ring-Diagram Analysis of Global Oscillation Network Group 
and Michelson Doppler Imager Data, 2006, {\bfseries 638}, 576--583

\bibitem[\protect\citeauthoryear{Howe} {2009}] {h09}
Howe, R., {\itshape Living Rev. Sol. Phys.}, Solar Interior Rotation 
and its Variation, 2009, {\bfseries 6}, no.1, 75pp

\bibitem[\protect\citeauthoryear{{Jouve et al}} {2008}] {jetal08}
Jouve, L., Brun, A.S., Arlt, R., Brandenburg, A., Dikpati, M., Bonanno, 
A., K\"apyl\"a, P.J., Moss, D., Rempel, M., Gilman, P., Korpi, M.J., and
Kosovichev, A.G., {\itshape Astron. Astrophys.}, 
A solar mean field dynamo benchmark, 2008, {\bfseries 483}, 949--960

\bibitem[\protect\citeauthoryear{K\"ohler} {1970}] {k70}
K\"ohler, H., {\itshape Sol. Phys.}, Differential Rotation Caused by 
Anisotropic Turbulent Viscosity, 1970, {\bfseries 13}, 3--18

\bibitem[\protect\citeauthoryear{Komm et al} {2012}] {kghbrh12}
Komm, R., Gonzalez-Hernandez, I., Hill, F., Bogart, R., Rabello-Soares, M.C.,
and Haber, D., {\itshape Solar Phys.}, Subsurface Meridional Flow from HMI
Using the Ring-Diagram Pipeline, 2012, {\bfseries}, DOI 10.1007/s11207-012-073-y

\bibitem[\protect\citeauthoryear{Miesch} {2005}]{m05}
Miesch, M.S., {\itshape Living Rev. Sol. Phys.}, Large-Scale Dynamics 
of the Convection Zone and Tachocline, 2005, {\bfseries 2}, no.1 pp

\bibitem[\protect\citeauthoryear{Rempel} {2005}] {r05}
Rempel, M., {\itshape The Astrophys. J.}, Solar Differential Rotation 
and Meridional Flow: The Role of a Subadiabatic Tachocline for the 
Taylor-Proudman Balance, 2005, {\bfseries 622}, 1320--1332 

\bibitem[\protect\citeauthoryear{R\"udiger} {1989}] {r89}
{R\"udiger}, G., {\itshape Akademie-Verlag, Berlin}, Differential 
Rotation and Stellar Convection, 1989, 328pp

\bibitem[\protect\citeauthoryear{{Rightmire-Upton}, {Hathaway} \& {Kosak}}
{2012}]{rhk12} Rightmire-Upton, L., Hathaway, D.H. \& Kosak, K., {\itshape
The Astrophys. J.}, Measurements of the Sun's High-Latitude Meridional
Circulation, 2012, {\bfseries 761}, L14--L17

\bibitem[\protect\citeauthoryear{{R\"udiger} \& {Hollerbach}} {2004}]{rh04}
R\"udiger, G., and Hollerbach, R., {\itshape Wiley, Weinheim}, The Magnetic 
Universe, 2004, 332pp.

\bibitem[\protect\citeauthoryear{{Skaley} \& {Stix}}{1991}]
{ss91} Skaley, D., and Stix, M., {\itshape Astron. Astrophys.},
The overshoot layer at the base of the solar convection zone, 1991, 
{\bfseries 241}, 227--232
  
\bibitem[\protect\citeauthoryear{{Svanda}, {Klvana} \& {Sobotka}}{2006}]
{sks06} Svanda, M., Klvana, M., and Sobotka, M., {\itshape Astron. Astrophys.},
Large-scale horizontal flows in the solar photosphere. I. Method and tests 
on synthetic data, 2006, {\bfseries 458}, 301--306
  
\bibitem[\protect\citeauthoryear{{Svanda}, {Kosovichev} \& {Zhao}}{2007}]
{skz07} Svanda, M., Kosovichev, A.G., and Zhao, J., {\itshape Sol. Phys.},
Comparison of Large-Scale Flows on the Sun Measured by Time-Distance 
Helioseismology and Local Correlation Tracking, 2007, {\bfseries 241}, 27--37
  
\bibitem[\protect\citeauthoryear{Tassoul} {2000}] {t00}
Tassoul, J. -L., {\itshape Cambridge Univ. Press}, Stellar Rotation, 
2000, 256pp

\bibitem[\protect\citeauthoryear{{Tassoul} \& {Tassoul}}{1995}]{tt95}
Tassoul, M. \& Tassoul, J.-L., {\itshape The Astrophys. J.}, 
Meridional Circulation in Rotating Stars. XI. Single-Cell Pattern versus 
Double-Cell Pattern, 1995, {\bfseries 440}, 789

\bibitem[\protect\citeauthoryear{Ulrich} {2010}] {u10}
Ulrich, R. K., {\itshape The Astrophys. J.}, Solar Meridional Circulation 
from Doppler Shifts of the Fe I Line at 5250 $A^o$ as Measured by the 150-foot
 Solar Tower Telescope at the Mt. Wilson Observatory, 2010, {\bfseries 725}, 
658--659

\bibitem[\protect\citeauthoryear{{Wang} \& {Sheeley}}{1991}]{ws91}
Wang, Y.-M. \& Sheeley, N.R.,Jr, {\itshape The Astrophys. J.}, Magnetic Flux
Transport and the Sun's Dipole Moment--New Twists to the Babcock-Leighton
Model, 1991, {\bfseries 375}, 761-770

\bibitem[\protect\citeauthoryear{Zhan et al} {2012}]{zsz12}
Zhan, X, Schubert, G., and Zhang, K., {\itshape Icarus}, Anelastic Convection
Driven Dynamo Benchmarks, A Finite Element Model, 2012, {\bfseries 218}, 345 

\end{thebibliography}
\end{document}